# A 12.5 GHz-Spaced Optical Frequency Comb Spanning >400 nm for near-Infrared Astronomical Spectrograph Calibration

F. Quinlan[1,*], G. Ycas[1,2], S. Osterman[3], S. A. Diddams[1,†]

[1]National Institute of Standards and Technology, 325 Broadway, 80305 Boulder, CO, USA
[2]University of Colorado Physics Dept., Boulder, CO USA
[3]University of Colorado, Center for Astrophysics and Space Astronomy, Boulder, CO, USA

*fquinlan@boulder.nist.gov, †sdiddams@boulder.nist.gov

**Abstract:** A 12.5 GHz-spaced optical frequency comb locked to a Global Positioning System disciplined oscillator for near-IR spectrograph calibration is presented. The comb is generated via filtering a 250 MHz-spaced comb. Subsequent nonlinear broadening of the 12.5 GHz comb extends the wavelength range to cover 1380 nm to 1820 nm, providing complete coverage over the H-band transmission window of Earth's atmosphere. Finite suppression of spurious sidemodes, optical linewidth and instability of the comb have been examined to estimate potential wavelength biases in spectrograph calibration. Sidemode suppression varies between 20 dB and 45 dB, and the optical linewidth is ~350 kHz at 1550 nm. The comb frequency uncertainty is bounded by +/- 30 kHz (corresponding to a radial velocity of +/-5 cm/s), limited by the Global Positioning System disciplined oscillator reference. These results indicate this comb can readily support radial velocity measurements below 1 m/s in the near-IR.

## I. INTRODUCTION

High precision astronomical spectroscopy in the near infrared (IR) permits lines of scientific inquiry unavailable in the visible and ultraviolet (UV) [1]. For example, the higher transmission in the IR through galactic and stellar dust allows the study of objects hidden in the visible and UV, such as young stars and planets still embedded in the clouds from which they formed. Also, the high red shift from distant universe objects requires IR spectroscopy to study spectra emitted in the visible and UV. Perhaps most exciting is extrasolar planet searches around M-dwarf stars via the radial velocity (RV) method. M-dwarfs are attractive candidates for extrasolar planet searches because they make up a large portion of the galactic neighborhood and their lower temperatures push the habitable zone closer to the star [2]. The closeness of the habitable zone, in conjunction with the smaller mass of these stars, creates larger RV signatures in a shorter time. Indeed, planet searches around M-dwarfs are a key component of the Exoplanet task force strategy [2].

The growing interest in near-IR high resolution (defined as $R = \lambda/\Delta\lambda$) astronomical spectroscopy is reflected in the number of spectrographs that have been under development in recent years, including CRIRES (R = 100,000) [1], Pathfinder (R = 50,000) [3], APOGEE (R = 20,000) [4] and FIRST (R = 50,000) [5]. A key component in realizing the potential of these spectrographs is precision calibration. Optical frequency combs from mode-locked lasers with multi-gigahertz spacing have recently been recognized as near ideal standards for the calibration of astronomical spectrographs [6].

The power of a frequency comb as a calibration source relies primarily on two properties of the comb.

First, a frequency comb consists of an array of discrete, regularly spaced frequency lines or modes spanning hundreds of nanometers. Second, the frequency of every comb mode is traceable to the SI second. The frequency of the *n*th mode of an optical frequency comb may be written as

$$v_n = n \times f_{rep} + f_{ceo}, \qquad (1)$$

where $f_{rep}$ is the repetition rate or mode spacing and $f_{ceo}$ is the carrier envelope offset frequency [7]. The spectrum of a frequency comb is fully determined once $f_{rep}$ and $f_{ceo}$ are known. Detection of $f_{rep}$ simply requires placing a fast photodiode in the beam path. Detecting $f_{ceo}$ is more involved—the most straightforward technique requires an octave-spanning spectrum and an f-2f interferometer [7]. Once detected, $f_{rep}$ and $f_{ceo}$ can be locked to oscillators traceable to the SI second, thus every mode of the frequency comb is also traceable to the SI second. Applying this "optical frequency ruler" to the calibration of an astronomical spectrograph would support RV measurements with cm/s precision and accuracy [6]. This represents more than an order of magnitude improvement over traditional calibration sources of emission lamps and absorption cells [8-10]. Moreover, the traceability of the spectrograph calibration allows comparisons from different instruments in different locations years or decades apart. Also, the density of lines from a frequency comb can allow compensation for systematic shifts due to inhomogeneity in the spectrograph that could otherwise limit RV precision [3, 11]. The high power per combline enables improvements in spectrograph slit illumination as well. Ideally one would deliver calibration light to the spectrograph via a uniform, isotropic source such as an integrating sphere. Whereas the loss ($10^{-6}$) of an integrating sphere is too great for traditional calibration sources, the high power per combline (100 nW or more) of the frequency comb permits the use of such a lossy coupling mechanism. A possible layout is shown in Fig. 1 [12].

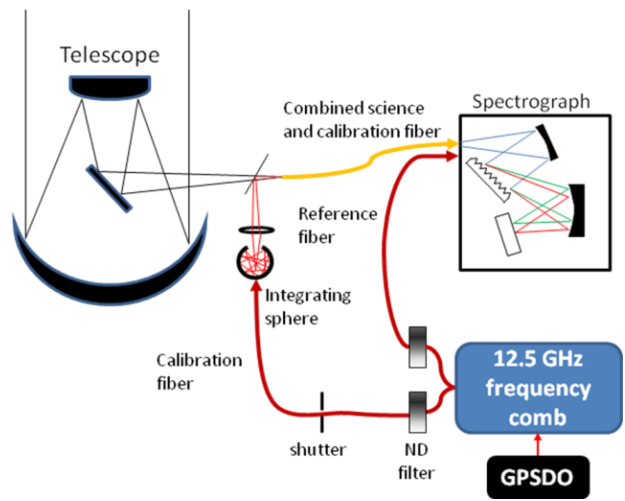

FIG. 1. Frequency comb calibration setup. Light from the frequency comb is split into two fibers—calibration and reference. The calibration fiber is fed to an integrating sphere for illumination of the spectrograph fiber. The reference fiber provides comb light spatially offset at the spectrograph detector array. ND, neutral density; GPSDO, Global Positioning System disciplined oscillator.

While frequency combs have the potential to make a significant impact, it is difficult to generate hundreds of nanometers spectral coverage with the requisite multi-gigahertz spacing directly from a mode-locked laser. Generally, the mode spacing from a frequency comb is limited to 1 GHz or less, requiring a wavelength resolution well above that of current spectrographs. For example, a resolution of

50,000 at 1550 nm would require a comb spacing of at least 12 GHz to ensure modes are separated by at least three resolution elements. A promising approach to make frequency combs available for spectrograph calibration is to thin a lower repetition rate comb with a filter cavity that transmits only a subset of modes [6, 11-19]. (It should be noted that the recent development of a 10 GHz Ti:Sapphire mode-locked laser spanning a wavelength range of 470 nm to 1130 nm is also a interesting possibility [20].) A filtered frequency comb has been used to simultaneously record the frequency comb and solar spectra onto an astronomical spectrograph with state of the art performance across ~18 nm, demonstrating the potential of this technique [11]. In this paper, we report on a 12.5 GHz-spaced optical frequency comb spanning over 400 nm for astronomical spectrograph calibration. The frequency comb is obtained via cavity filtering the output of a lower repetition rate mode-locked laser. Spectral broadening of the 12.5 GHz-spaced comb in nonlinear optical fiber yields coverage from 1380 nm to 1820 nm, supplying calibration lines across the H-band (1500 nm to 1800 nm). The mode filtering technique can introduce small biases into spectrograph calibration, the magnitude of which depends on the implementation. Design considerations leading to our implementation and possible wavelength biases resulting from our setup are discussed. This discussion provides motivation for measurements made on the optical spectrum in terms of frequency accuracy and stability, optical linewidth, and suppression of spurious sidemodes. To our knowledge, the data presented here represent the most comprehensive characterization of an optical frequency comb for spectrograph calibration reported to date.

## II. 12.5 GHZ-SPACED COMB GENERATION

Generation of the 12.5 GHz-spaced optical frequency comb starts with a 250 MHz repetition rate passively mode-locked Erbium-doped fiber laser. Part of the laser output is amplified, then passed through highly nonlinear fiber (HNLF) to generate an octave-spanning spectrum. The broadened optical spectrum, spanning 1 μm-2 μm, is sent to a standard f-2f interferometer for $f_{ceo}$ detection [7]. A separate laser output is used for repetition rate detection. The carrier envelope offset frequency and $f_{rep}$ are both locked to low noise synthesizers that are referenced to a Global Positioning System disciplined oscillator (GPSDO). Both $f_{ceo}$ and $f_{rep}$ stay locked for several days without intervention.

With both $f_{rep}$ and $f_{ceo}$ locked, the frequency of the $n$th mode is given by

$v_n = n \times 250$ MHz $- 116$ MHz. (2)

For $v_n$ near 1550 nm, $n \sim 773,000$. The $f_{ceo}$ value of 116 MHz was chosen for best transmission through the mode selection filter cavity [21]. A schematic of the 250 MHz frequency comb is shown in Fig 2.

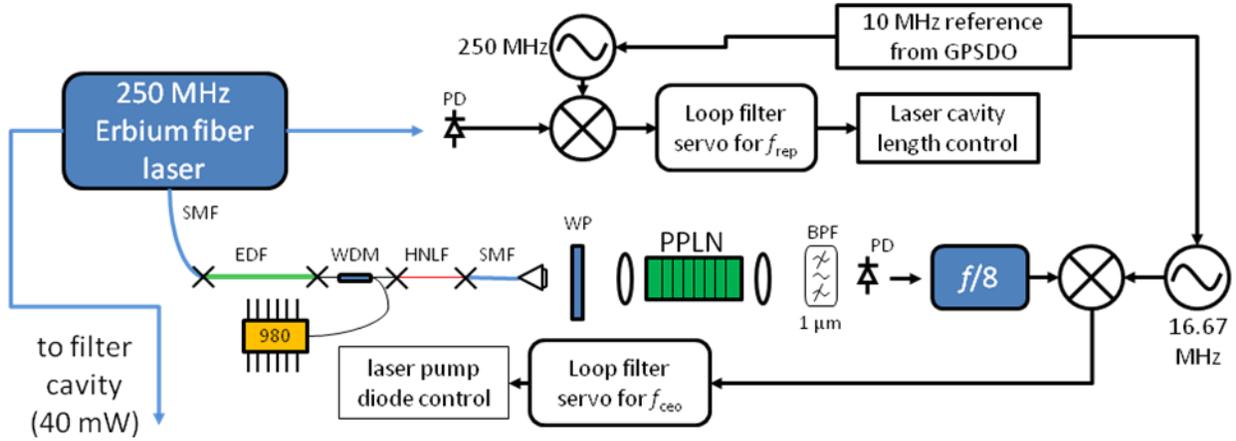

FIG. 2. 250 MHz-spaced optical frequency comb source. $f_{rep}$ is controlled by changes to the laser cavity length, while $f_{ceo}$ is controlled by changes to the laser pump power. Both $f_{rep}$ and $f_{ceo}$ are locked to oscillators referenced to a GPSDO. $f_{ceo}$ is locked via the copy of $f_{ceo}$ at $f_{rep} - f_{ceo}$ (134 MHz). For the $f_{ceo}$ lock, part of the laser output is amplified with Erbium-doped fiber (EDF) before the highly nonlinear fiber (HNLF) that broadens the spectrum to cover 1 μm - 2 μm. Light at 2 μm is frequency-doubled in periodically-poled Lithium Niobate (PPLN). Light at 1 μm passes through the PPLN and interferes on the photodetector (PD) with the frequency-doubled light. BPF, bandpass filter; SMF, single-mode fiber; WDM, wavelength division multiplexer; WP, waveplate.

Another laser output of 40 mW is sent to the first of two mode selection filter cavities. Both filter cavities are high finesse (2000 at 1550 nm) air-gap etalons with a free spectral range of 12.5 GHz. The periodic transmission of the filter cavities transmits modes separated by 12.5 GHz (one mode out of every 50) with low loss, while other modes are greatly suppressed. Both filter cavities are locked to the frequency comb by imparting a ~50 kHz dither on the filter cavity length and locking to a peak in the transmitted power. This lock has also shown to be robust for several days of continuous operation.

After the first cavity, the laser is sent through a system of amplifiers and dispersion compensating fiber. Amplification and pulse compression are necessary to raise the peak power of the pulse train for nonlinear broadening later on. The first two amplifiers are semiconductor optical amplifiers (SOAs) with gain peaks offset by ~40 nm. This arrangement preserves the 50 nm of bandwidth after the first filter cavity and raises the average power of the pulse train from 200 μW to 80 mW. This level is sufficient to seed the high-power Erbium-doped fiber amplifier (HPEDFA) that follows. The average power after the HPEDFA is ~1.4 W.

After amplification, the pulse train passes through the second filter cavity. The frequency comb is then broadened in 50 m of low dispersion HNLF [22] to generate over 400 nm of 12.5 GHz-spaced modes. The HNLF has a dispersion slope of +0.019 $ps^2$/km/nm, a nonlinear coefficient of 30/W/km, and the zero dispersion wavelength is 1550 nm. Losses coupling into and out of the second filter cavity as well as insertion loss in the isolator following the HPEDFA result in ~400 mW into the HNLF. The pulse width at the HNLF input is ~300 fs, resulting in a peak power of 100 W. The mode-locked laser and the filter cavities rest on a vibration isolation optical table and are covered with a foamboard box to provide a layer of vibration, acoustic and thermal isolation. A schematic showing the generation of the 12.5 GHz-spaced comb is shown in Fig. 3.

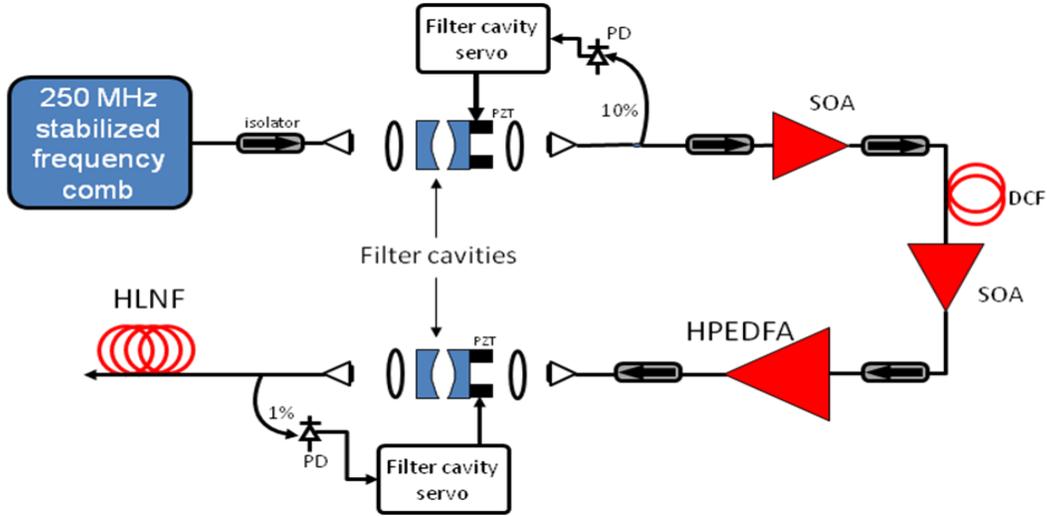

FIG. 3. 12.5 GHz-spaced comb generation. HPEDFA, high power Erbium-doped fiber amplifier; SOA, semiconductor optical amplifier; PZT, piezoelectric transducer; DCF, dispersion compensating fiber. All fibers are single mode at 1550 nm.

## III. DESIGN CONSIDERATIONS AND WAVELENGTH BIASES

There are important limitations that must be considered when one uses a frequency comb with cavity filtering for spectrograph calibration that warrant careful measurements of the resulting comb spectrum [6, 13, 14]. Here we discuss the potential limitations of our particular realization of a 12.5 GHz-spaced comb and how we have worked to mitigate these limitations.

### A. Higher order spatial modes, filter cavity dispersion and finesse

Higher order spatial modes (HOMs) of the filter cavity can be excited by the comb, resulting in unwanted laser modes transmitted with low loss [23]. A judicious choice of curvatures in the etalon mirrors can mitigate this effect [13]. For the etalons presented here, there are HOMs with a frequency offset of +1.75 GHz from the main 12.5 GHz spaced modes that are transmitted with low loss. Fortunately the overlap between the HOM and the optical fiber spatial mode is small, and no 1.75 GHz-offset frequency component is detectable after the second filter cavity to the -70 dB level. A larger problem stems from dispersion in the filter cavity. Whereas the modes of the mode-locked laser are equally spaced, dispersion in the cavity mirrors results in unequal spacing of the filter cavity transmission peaks. This causes a walk-off between the laser modes and the etalon transmission peaks, limiting the bandwidth of the filtered comb [6, 13, 14]. One could increase the useable bandwidth by decreasing the finesse, but this comes at the expense of reduced sidemode suppression. For the system presented here, we have chosen a finesse of 2000, yielding, in a single pass, ~38 dB suppression of the nearest neighbor modes and 50 nm bandwidth.

## B. Filter cavity induced wavelength biases

In addition to limiting the optical bandwidth, the shift between a laser mode and the transmission peak of the etalon can result in an apparent wavelength shift in two ways. First, the laser modes are of finite width and will be asymmetrically reshaped as the mode is detuned from the transmission peak of the filter cavity. For detunings such that the mode remains within the full width at half maximum (FWHM) of the transmission peak, as long as the linewidth is less that 1/10 of the FWHM of the etalon, the wavelength shift should be less than 1 cm/s [12]. Second, the shift between the laser mode and the center of the transmission peak of the etalon will result in asymmetric suppression of the 250 MHz-offset modes [14]. The unequal weight of these sidemodes, unresolved by the spectrograph, will also cause a shift in the line center. For example, consider an R = 50,000 spectrograph with a Gaussian transfer function that is illuminated with a combline that has asymmetrically suppressed modes offset by 250 MHz. Neglecting detector noise, for sidemodes suppressed 20 +/-1 dB, 30 +/-1 dB and 40 +/-1 dB, the estimated RV bias from the 250 MHz-offset modes is 1.8 m/s, 0.18 m/s and 0.018 m/s, respectively [14]. Note that even if shifts in the observed wavelength caused by filtering are stable for long periods, it weakens one of the major advantages of the comb, namely traceability. A frequency comb calibration that is tied to a particular filter cavity configuration complicates comparisons with different comb sources calibrating different spectrographs.

## C. Spectral broadening in HNLF

Bandwidth limitation of the filter cavities motivated our use of HNLF to spectrally broaden the 12.5 GHz-spaced comb to provide full H-band coverage. Nonlinear spectral broadening of a high repetition rate pulse train comes with its own set of challenges. Because the available energy is divided into more pulses, nonlinear spectral broadening requires a higher average power of the pulse train to achieve the same peak power as a lower repetition rate source. For example, consider the 250 MHz pulse train used for the f-2f interferometer and $f_{ceo}$ detection described above. To achieve an octave of bandwidth in 50 cm of fiber, the average power of the pulse train is 100 mW. For the 12.5 GHz pulse train to have the same peak power (assuming the pulse width is the same), the average power would need to be 50 times higher, or 5 W. Achieving this level of output power requires amplification, but amplifying our pulse train to several watts without spectral gain narrowing (which would broaden the pulse) or detrimental nonlinear effects in the amplifier would be extremely challenging. Also of concern is increased noise as the amplified comb is passed through the HNLF. The intensity noise of a pulse train is known to increase as it passes through HNLF [24, 25]. In particular, the amplified spontaneous emission (ASE) from the amplifiers will degrade the coherence of the optical spectrum [26, 27]. Also, four-wave-mixing (FWM) between the 12.5 GHz-spaced modes and the suppressed 250 MHz-spaced modes can significantly reduce the suppression ratio. We have attempted to mitigate these effects by placing a second filter cavity just before the HNLF. The second filter cavity removes ASE between the comb modes and increases the suppression of the 250 MHz offset sidemodes.

## D. Accuracy and stability of reference oscillator

Finally we note that the accuracy and stability of a frequency comb are ultimately determined by its frequency reference. We have chosen a GPSDO as the frequency reference for our frequency comb as it provides transportable, long term accuracy at a reasonable cost. While the GPSDO frequency is expected to track UTC(USNO) in the long term (accuracy better than $10^{-12}$ in a day), short term fluctuations on the order of a few parts in $10^{-10}$ are possible. The performance among different GPSDOs can vary significantly [28], therefore measurements of the GPSDO are warranted, as well as confirming the comb tracks it after the filter cavities, amplifiers and HNLF.

As these concerns indicate, the comb must be measured in detail to confirm spectrograph calibration at the cm/s level. In the next section we report on measurements of 250 MHz-offset sidemode suppression, optical frequency accuracy and stability, and optical linewidth.

# IV. RESULTS

## *A.* Optical spectrum and sidemode suppression

The optical spectra for various points along the lightpath are shown in Fig. 4. Immediately after the first filter cavity, a reduction in the wings of the spectrum is evident—a result of the walk-off between the filter cavity transmission peaks and the laser modes. At the center of the spectrum, the loss from coupling to the filter cavity and single mode fiber is ~4dB. The optical spectrum after the SOAs maintains the spectral FWHM while the power per mode increases more than a factor of 100. The spectrum after the second filter cavity is also shown. Note the bandwidth reduction due to gain narrowing in the HPEDFA. This represents the optical spectrum of the pulse train into the HNLF. The optical spectrum following the HNLF is shown in Fig. 5. The comb extends from 1380 nm to 1820 nm, spanning the H-band. From 1300 nm to 1700 nm, the spectrum was measured with a high resolution (R = 77,000) optical spectrum analyzer. For wavelengths longer than 1700 nm, the spectrum was measured with a much lower resolution monochromator, thus only the spectral envelope is available for λ > 1700 nm. Shorter lengths of the same HNLF were also tested where it was observed that the peak near 1820 nm moved to shorter wavelengths. This is consistent with a soliton self-frequency shift [29]. The power of the 12.5 GHz-spaced comblines varies about 15 dB from 10 μW per mode to 320 μW per mode.

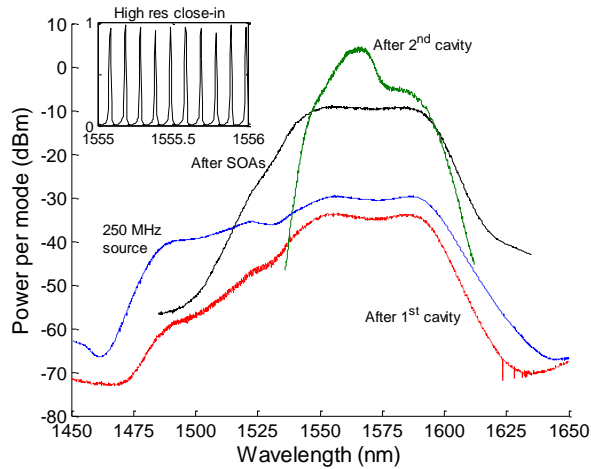

FIG. 4. Power per optical mode at different stages of comb filtering and amplification. The inset is a high resolution close-in view of the spectrum after the SOAs.

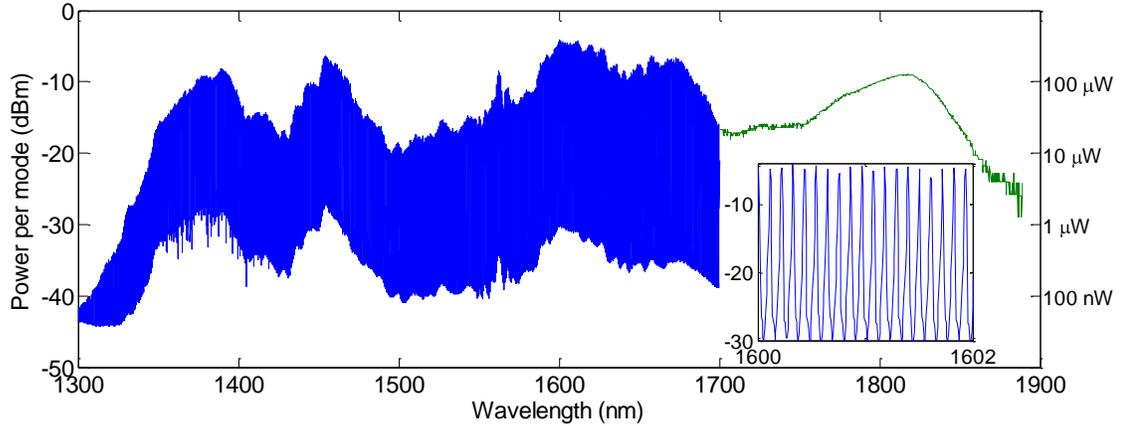

FIG. 5. Power per optical mode after nonlinear broadening. The inset is a high resolution close-in view around 1600 nm. The comb signal-to-noise is limited by the grating-based optical spectrum analyzer employed for this measurement. The asymmetry of the modes in the inset is also due to the optical spectrum analyzer.

The level of suppression of the 250 MHz-offset modes, asymmetry in mode suppression and how mode suppression varies with wavelength were measured. These measurements were performed by optically heterodyning the 12.5 GHz-spaced comb with a tunable continuous-wave (CW) laser. The microwave spectrum of the photodetected beat between the CW laser and the frequency comb directly gives the optical sidemode suppression. An example microwave spectrum trace of a CW-comb beat is shown in Fig. 6(a). Two tunable lasers, spanning 1380 nm to 1490 nm and 1520 nm to 1630 nm, respectively, were tuned across the comb spectra in 10 nm steps. For measurements on the comb after the HNLF, the measurement span was limited by the tuning range of our tunable lasers. The results of all the beats measured are summarized in Fig. 6(b). Immediately after the first filter cavity, the suppression is ~38 dB in the center of the spectrum and ~31 dB at 1525 nm. This is consistent with the reflectively of the mirror coatings: 99.89 % from 1550 nm to 1600 nm but dropping to ~99.85 % at 1510 nm. After the SOAs, the suppression is reduced considerably, likely due to FWM in the SOAs. Also note the asymmetry as great as 7 dB in the sidemodes, particularly at 1580 nm and 1590 nm. The second filter cavity increases the suppression of the sidemodes to around 60 dB and no asymmetry was detected within the measurement uncertainty of +/-1 dB. The level of suppression changes again after the HNLF, ranging from 20 dB at 1380 nm to 45 dB at 1570 nm. Again, no asymmetry in the sidemode strength was observed within +/-1 dB.

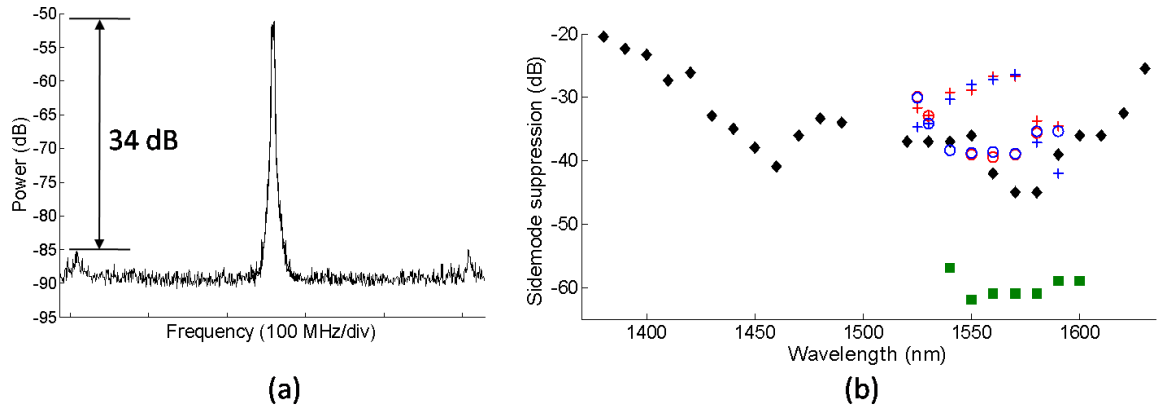

FIG. 6. 250 MHz-offset mode suppression measurements. (a) Example heterodyne measurement where the higher and lower frequency sidemodes are suppressed 34 dB. (b) Summary of all sidemode suppression measurements. Circles: after the 1st filter cavity, showing higher (blue) and lower(red) frequency sidemode suppression. Crosses: after the SOAs, showing higher and lower frequency sidemode suppression. Squares: after the second filter cavity, where no asymmetry in sidemodes was detected. Black diamonds: after HNLF where no asymmetry was detected.

Ideally, the sidemode suppression on every 12.5 GHz-spaced combline would be measured. From 1380 to 1630 nm the measurements are representative, for we consider it unlikely that the suppression ratio changes more than a few dB in 10 nm. Less certain is the sidemode suppression beyond 1630 nm. Other possible techniques to measure suppression on every 12.5 GHz-spaced combline over the entire wavelength range are two comb spectroscopy [30], or a high resolution Fourier transform spectrometer. The use of these measurement techniques is currently under investigation.

Another method that has been reported in the literature to calculate the sidemode suppression ratio is to photodetect the filtered comb and measure the sidemode strength in the photodetected microwave spectrum [13, 18, 19]. We find this method unreliable for the following reasons. First, no information about asymmetry of sidemode suppression or variations of suppression with wavelength can be obtained. Second, using the photodetected pulse train to measure optical sidemode suppression requires assumptions about the relative phase among the optical modes. As an example of how such assumptions can be misleading, consider the photodetected spectrum of the pulse train after the SOAs, shown in Fig. 7(a). This spectrum shows 250 MHz offset spurs are >57 dB below the 12.5 GHz carrier. If we make the assumption that all modes are in phase, we would predict an optical sidemode suppression of 63 dB. However, as shown in Fig. 7(b), an optical heterodyne measurement at 1560 nm reveals the optical sidemode suppression after the SOAs of 27 dB.

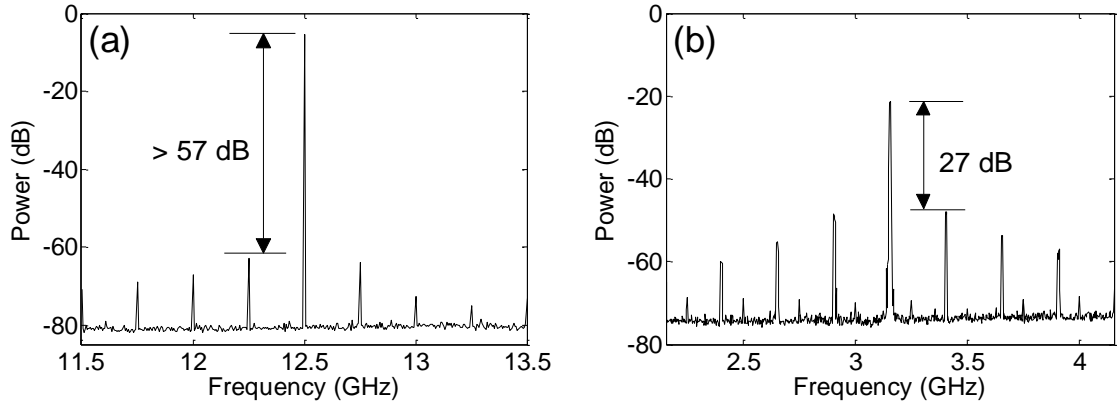

FIG. 7. (a) Microwave spectrum of the photodetected pulse train after the SOAs. Sidemodes offset by 250 MHz from the 12.5 GHz carrier are 57 dB below the carrier. (b) optical heterodyne measurement at 1560 nm where only 27 dB suppression is observed. This heterodyne measurement is also represented in Fig. 6.

## B. Frequency accuracy and stability

Frequency accuracy and stability of the 12.5 GHz-spaced comb were measured. These measurements made use of frequency counters referenced to a 10 MHz Hydrogen-Maser signal, traceable to the SI second, with stability near $1 \cdot 10^{-13}$ in one second. For all measurements, the frequency counter gate time was one second. The 10 MHz reference from the GPSDO, $f_{rep}$, $f_{ceo}$ and a comb mode at 1550 nm ($n$ = 773,587) were simultaneously measured. Measurements were made along the lightpath to see the effects, if any, of the filter cavities, amplifiers and HNLF.

Because the frequency of the mode near 1550 nm is not directly countable, a separate reference frequency comb was used to generate a beat tone that could be counted. The setup for generating the beat tone between the reference comb and the mode near 1550 nm is shown in Fig. 8. The reference comb is an octave-spanning, 1 GHz repetition rate Ti:sapphire laser with both $f_{rep}$ and $f_{ceo}$ frequencies referenced to the H-Maser [31]. A CW fiber laser at 1550 nm is frequency-doubled to 775 nm and beat against the reference comb. The beat signal between the reference comb and the frequency-doubled fiber laser is then used to lock the frequency of the fiber laser to the reference comb. A beat signal between the fiber laser output at 1550 nm and the comb for spectrograph calibration is then photodetected and counted. In this way the stability and accuracy of a mode near 1550 nm can be measured against the H-Maser.

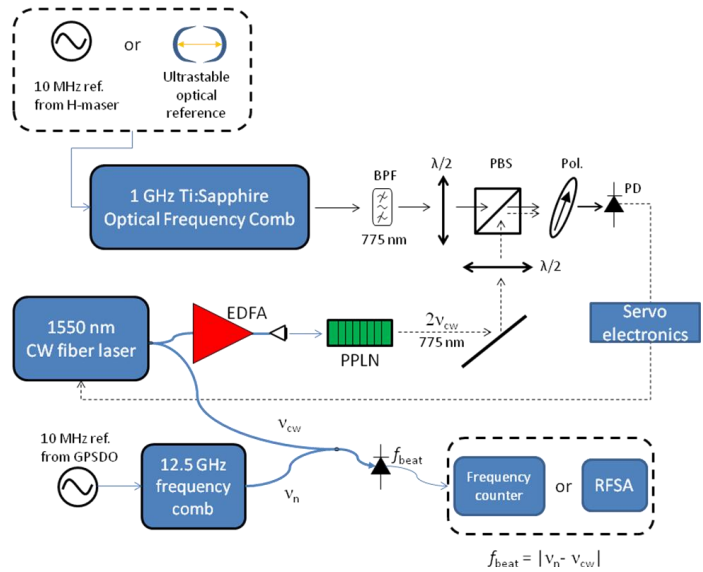

FIG 8. Setup for measuring the frequency stability, accuracy and linewidth of an optical mode. Light from the Ti:Sapphire frequency comb is combined with the frequency-doubled light from the CW laser via a polarization beam splitter (PBS) and polarizer (Pol.). Half-wave plates (λ/2) control the polarization into the PBS. For frequency accuracy and stability measurements, the 1 GHz Ti:Sapphire reference comb is locked to the H-Maser and the beat frequency, $f_{beat}$, is sent to the frequency counter. For linewidth measurements the reference comb is locked to an ultrastable optical reference and $f_{beat}$ is sent to an rf spectrum analyzer (RFSA). EDFA, erbium-doped fiber amplifier.

From the frequency counter record, the fractional frequency offset was determined by subtracting the nominal frequency, as determined by the H-Maser, and normalizing the frequency to its carrier. Fractional frequency offset measurements for GPSDO, $f_{rep}$, $f_{ceo}$ and a comb mode after the HNLF are shown in Fig. 9. Note the fractional frequency of $f_{rep}$, $f_{ceo}$ and the comb mode at 1550 nm all track the GPSDO reference and are on the order of a few parts in $10^{-10}$. Since the fractional frequency instability is clearly dominated by the GPSDO reference, it is interesting to examine the difference in the fractional frequency offset of the comb mode [Fig. 9(d)] and the GPSDO [Fig. 9(a)]. As shown in Fig. 10, this difference is bounded by +/- 1 $10^{-11}$ with a standard deviation 6.4 $10^{-13}$, indicating significantly improved performance should be possible with a better reference oscillator.

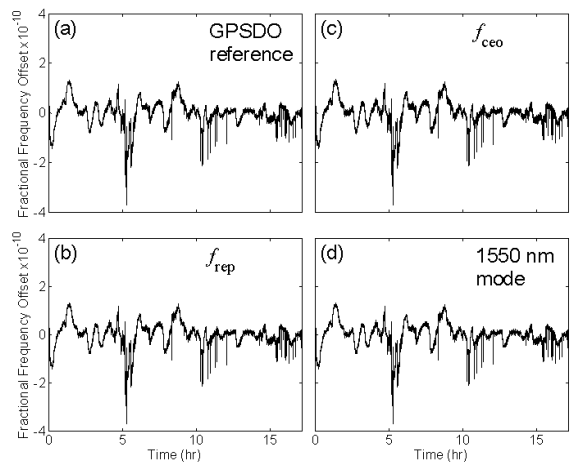

FIG. 9. Fractional frequency offset of (a) the GPSDO, (b) fceo, (c) frep and (d) an optical mode near 1550 nm.

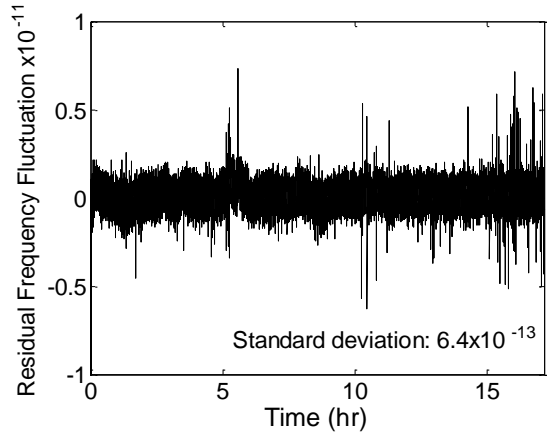

FIG. 10. Residual fractional frequency offset of the 1550 nm mode relative to the GPSDO.

Similar measurements were made to examine the comb mode directly from the 250 MHz laser, after the first filter cavity, and after the second filter cavity. In all cases the optical mode tracked the GPSDO as in Fig. 9. Fig. 11 shows the stability and accuracy of the comb mode in both frequency and RV units along the lightpath. For all measurements, obtained over a one week period, RV offsets are mostly bound between +/- 5 cm/s. There are a few momentary (less than 3 s duration) excursions approaching 10 cm/s originating from the GPSDO. Depending upon the integration time of the spectrograph detector array, these short term excursions may be averaged over. Note that these measurements only track the instability of one of the 12.5 GHz-spaced modes—any constant offset due to asymmetry in suppressed sidemodes is not taken into account.

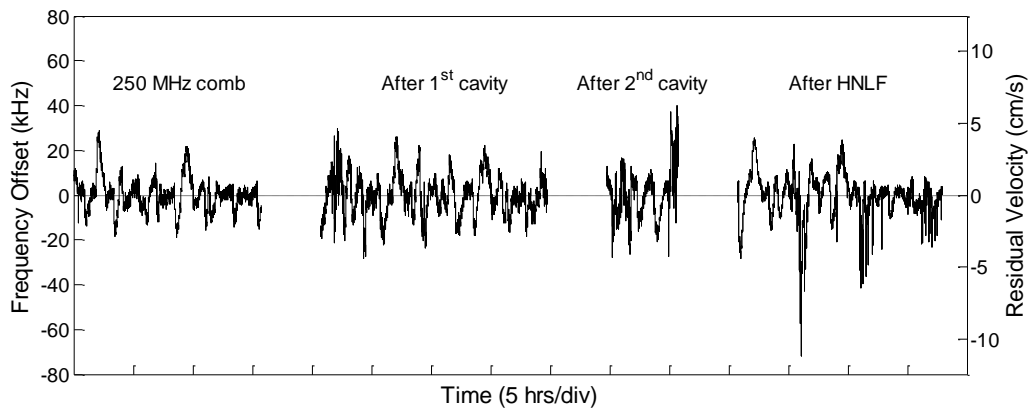

FIG. 11. Frequency and radial velocity bias of an optical mode near 1550 nm measured along the lightpath. Zero offset is determined by the H-Maser. For the measurement after the HNLF, the mode number *n* and frequency were determined to be 773587 and 193.9966 THz, respectively. For all measurements the accuracy and stability is dominated by the GSPDO.

## C. Optical linewidth

As noted above, a broad optical linewidth can introduce a bias into spectrograph calibration if the mode is asymmetrically shaped by one of the filter cavities. Because the finesse of 2000 for our filter cavities corresponds to a transmission peak FWHM of 6.25 MHz, the optical linewidth into the filter cavities should be less than 625 kHz to avoid any significant wavelength biasing due to line reshaping. The linewidth from a mode-locked laser comb source is largely determined by the short term stability (<1 s) of its reference, although the free running linewidth, and the $f_{rep}$ and $f_{ceo}$ locking electronics also have an effect. As shown above, it is the fractional frequency noise of the reference that will be transferred to the comblines, therefore the noise of the reference signal must be multiplied by the frequency ratio of the combline to the reference. Thus a mode near 1550 nm can have a linewidth of 100s of kHz when the 10 MHz reference signal linewidth is sub-Hertz. Aside from the 10 MHz reference, the components used to generate the 12.5 GHz comb can also affect the linewidth, most notably the HNLF. The optical linewidth for a mode near 1550 nm was measured along the lightpath. The setup for measuring the optical linewidth is nearly identical to that of the frequency accuracy and stability measurements. The only difference is the H-Maser frequency reference is replaced with a sub-Hertz linewidth optical reference [32]. The stability of the optical reference is transferred to the reference comb and the 1550 nm CW laser [33]. The heterodyne beat between the CW laser and the spectrograph comb was measured with a microwave spectrum analyzer to determine the linewidth. The optical linewidth at various points along the lightpath is shown in Fig. 12. Fig. 12(a) shows the linewidth measurement on a linear scale with a 1 MHz span. There is no significant change in the ~340 kHz FWHM linewidth as the comb passes through the filter cavities and the HNLF. The linewidth differences are more apparent in Fig. 12(b), where the intensity is plotted on a log scale and the span increased to 50 MHz. Note the high noise floor of the measurement after the HPEDFA, resulting from ASE from the amplifiers. This floor reduces dramatically as the ASE is filtered by the second filter cavity. After the HNLF, skirts appear staring 25 dB below the peak and drop to 60 dB below the peak at 25 MHz offset. The origins of these skirts could be the nonlinear increase of technical intensity noise on the pulse train and ASE that survives the second filter cavity parametrically mixing with the comb [22, 24-26].

Although these measurements were performed on only one comb mode, they do provide information on the linewidth of other modes. The linewidth varies with frequency, but for a spectrum spanning less than an octave, the linewidth will vary by less than a factor of 2 across the spectrum. Since the wavelength band through the filter cavities is narrow, the linewidths of the modes passing through the filter cavities are relatively constant at 340 kHz. This is below our 625 kHz criterion, therefore wavelength bias due to line reshaping should not be a concern. Again we note that with a better reference oscillator the linewidth can be reduced. After the HNLF, the linewidth FWHM should vary according to the mode frequency. Further investigation is required to determine how the magnitude of the skirts varies across the spectrum.

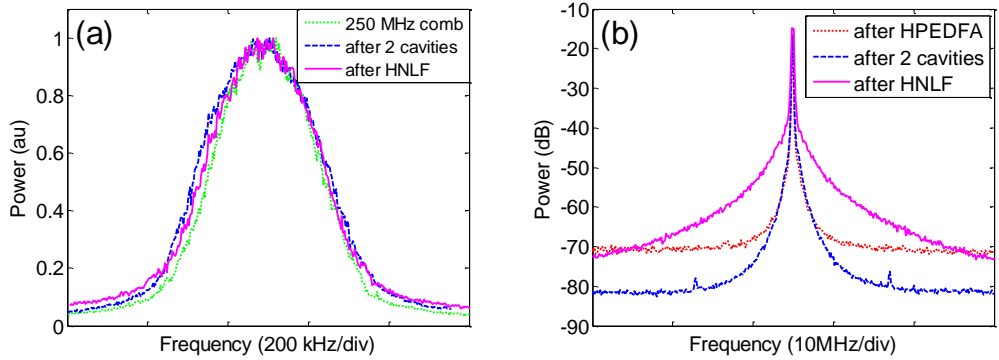

FIG. 12. Linewidth of an optical mode near 1550 nm. (a) Linear scale measurement of the linewidth along the lightpath. For each trace the spectrum analyzer sweep time is 5 msec and averaged 100 times. (b) Log scale measurement of the linewidth after the HPEDFA, after the second filter cavity and after the HNLF. For each trace the spectrum analyzer sweep time is 113 msec and averaged 100 times. Note the difference in frequency scales for (a) and (b).

## V. CONCLUSION

With growing interest in near-IR high resolution astronomical spectroscopy, precise spectrograph calibration becomes vital. To this end, a 12.5 GHz-spaced optical frequency comb spanning 1380 nm to 1820 nm has been built and characterized. The comb spacing was chosen for R = 50,000 spectrograph, placing one combline every 3 resolution elements. Measurements on sidemode suppression, optical mode frequency accuracy, stability and linewidth indicate sub-m/s calibration should be possible across the H-band. The frequency stability of the frequency comb is currently limited by the GPSDO to ~ +/- 5 cm/s. Higher performance GPSDOs are commercially available, therefore improvements on the comb frequency stability are likely. If calibration across a smaller wavelength span is required, the setup can be simplified by removing the HPEDFA and the HNLF. Measurements on this simplified setup demonstrated 60 dB sidemode suppression in a 50 nm FWHM bandwidth and more than -17 dBm (20 µW) per mode. Although the comb developed here is targeted for R = 50,000 in the H-band, the techniques used to characterize the comb should be applicable to frequency combs developed for other wavelength bands and other spectrograph resolutions as well.

In the hunt for exoplanets, precision calibration with a frequency comb is only one element necessary to measure a star's radial velocity with cm/s precision. Intrinsic limitations such as density of features in the stellar spectrum and stellar rotations, as well technical limitations such as signal-to-noise ratio on the spectrograph detector and pointing stability of the telescope will all affect the final RV precision [10]. However, spectrograph calibration is currently the dominating uncertainty, limiting RV precision in the H-band to ~10 m/s [3, 10]. Calibrating the spectrograph with a frequency comb should help facilitate RV precision limited by the intrinsic properties of the starlight.

# ACKNOWLEDGMENTS

We thank T. Fortier for the use of the Ti:sapphire reference comb, Y. Jiang for supplying the sub-Hertz optical reference, Y. Jiang and D. Braje for contributions in building the 250 MHz comb source, M. Lombardi and A. Novick for assistance with the GPSDO, and M. Hirano of Sumitomo Electric Industries for use of the HNLF. Financial support is provided by NIST and the NSF. F. Quinlan is supported as an NRC/NAS postdoctoral fellow. This work is a contribution of an agency of the US government and is not subject to copyright in the US.